\documentclass[12pt]{iopart}
\usepackage{graphicx}
\usepackage{cite}

\begin{document}

\title[A comprehensive study of decay modes associated with Pb isotopes]{A comprehensive study of decay modes associated with Pb isotopes}

\author{R. Sharma$^{1, 2}$, A. Jain$^{1,3}$, P. K. Sharma$^4$, S. K. Jain$^1$, and G. Saxena$^3$ }

\address{$^1$ Department of Physics, School of Basic Sciences, Manipal University Jaipur-303007, India}
\address{$^2$ Department of Physics, S. S. Jain Subodh P.G.(Autonomous) College, Jaipur-302004, India}
\address{$^3$ Department of Physics (H$\&$S), Govt. Women Engineering College,Ajmer - 305002, India}
\address{$^4$ Govt. Polytechnic College, Rajsamand-313324, India}
\ead{gauravphy@gmail.com}

\vspace{10pt}
\begin{indented}
\item[]December 2021
\end{indented}

\begin{abstract}
Decay modes in Pb isotopes within the range 176$\leq$A$\leq$266 are investigated by the calculation of half-lives using several empirical formulas. These formulas along with various theoretical treatments are first tested to reproduce experimental half-lives and known decay modes of Pb isotopes, which are in consequence applied to estimate half-lives and decay modes of unknown Pb isotopes. A comparison between $\alpha$-decay and weak-decay from the stable to drip-line isotopes is canvassed which leads to the excellent match with experimental data and applicability of applied empirical formulas. In addition, the full chain of Pb isotopes is probed as potential daughter candidates of cluster emission from superheavy nuclei which ensued the predominant role of $^{208}$Pb and nearby isotopes in probable cluster emission.

\end{abstract}
\noindent{\it keywords}: Alpha Decay; Weak-decay; Cluster Decay; Half-lives.
%\maketitle
\section{Introduction}
Nuclei with the heaviest well-established proton shell closure (Z$=$82) i.e. Pb isotopes, out of which 43 are experimentally accessible, offer an excellent ground for testing the theories and their continuous growths. The most neutron-rich lead isotope identified experimentally is $^{220}$Pb \cite{alvarez2010} whereas $^{178}$Pb is the most neutron-deficient lead isotope known recently \cite{badran2016}. On the other hand, $^{204,206,207,208}$Pb are stable and rest of the isotopes decay via $\alpha$, $\beta^{+}/EC$ or $\beta^{-}$ emissions. In addition, doubly magicity and core shape of $^{208}$Pb make it a predominant daughter nucleus of various cluster emissions as well as the terminating nucleus of several $\alpha$-decay chains of superheavy nuclei. Hence, interconnection among various decays in Pb isotopes classifies it uniquely for probing a variety of decay modes at the same time.\par

Recently, $\alpha$-decay has been contemplated in Pb isotopes using Coulomb and proximity potential model (CPPM) \cite{santosh2015}, the quantum-mechanical
tunneling mechanism of penetration through a potential barrier \cite{hosseini2019}, generalized liquid drop model \cite{liu2020} and modified generalized liquid drop model (MGLDM) \cite{santosh2020}. Additionally, $\beta$-decay half-lives are reported for $^{215 - 218}$Pb beyond N$=$126 by R. Caballero-Folch \textit{et al.} \cite{folch2017}. Another decay mode which was first proposed theoretically in 1980 by Sandulescu \textit{et al.} \cite{Sandulescu} before its experimental realization in 1984 by Rose and Jones \cite{Rose}, and intimately connected to closed-shell daughters as $^{208}$Pb or its neighbors, is the emission of heavy nuclei (clusters), such as carbon, oxygen, fluorine, neon, magnesium, silicon, etc. Therefore, the study of Pb isotopes which are either stable or decay via $\alpha$, $\beta^{+}/EC$, $\beta^{-}$, along with their potential as daughter nuclei of cluster emission in superheavy nuclei, provide plenteousness base for testing and improvements of theoretical approaches or empirical formulas, which has invoked us for the present study. \par

In this study, various theoretical models and empirical formulas are examined to produce known half-lives of Pb isotopes (176$\leq A\leq$266) considering $\alpha$-decay as good as the weak-decay. Consequently, these formulas are used to estimate half-lives of unknown Pb isotopes. In addition, half-lives for various cluster emissions considering Pb isotopes as daughter nuclei are examined to speculate various cluster emissions in the superheavy regions.
%%%%%%%%%%%%%%%%%%%%%%%%%%%%%%%%%%%%%%%%%%
\section{Theoretical Frameworks} \label{section-theory}
As mentioned above, decay modes of Pb isotopes within the range 176$\leq A\leq$266 are investigated by evaluating the decay half-lives using empirical formulas that are predominantly dependent on the energy released in the specific transition. The energy released in $\alpha$-decay, $\beta^-$-decay, and electron capture (EC) are denoted by $Q_\alpha$, $Q_{\beta^-}$, and $Q_{EC}$, respectively, and are estimated using a few widely used theories wherever the experimental values of these energies are not available. To evaluate the Q-values in the unknown regions, we have used various parameterizations/variants of the RMF models that have been practiced to a great extent. The commonly used variants are NL3$^*$ \cite{Lalazissis09}, TMA \cite{sugaTMA}, FSU-Gold \cite{Todd-Rutel05}, FSU-Garnet \cite{Chen15}. The relativistic mean-field approach and its variants are adequately described in various Refs. \cite{Todd-Rutel05,Serot1984,Ring1996,Lalazissis97,Lalazissis05,Niksic08,saxenajpg,saxena21} and our work \cite{Sharma21} which can be referred for further details of the formalism. For comparison, the results of other theoretical models viz. RCHB \cite{RCHB}, FRDM \cite{frdm2012} and WS4 \cite{ws4} are also taken into account. Aforesaid models are tested to reproduce the experimental Q-values \cite{audi20202} of Pb isotopes. Hence, we have calculated Q$_\alpha$, $Q_{\beta^-}$, and $Q_{EC}$ (respective equations are provided in the upcoming sections) values for approx 43 Pb isotopes in the range 96$\leq$N$\leq$138. The root-mean-square error RMSE (which provides error in the estimated data) along with standard deviation $\sigma$ (which gives deviation from experimental data) are given by the following equations, respectively, and are listed in Table \ref{rmse}.
\begin{eqnarray}\label{rmse1}
 RMSE &=& \sqrt{\frac{1}{N_{nucl}}\sum^{N_{nucl}}_{i=1}\left({x}\right)^2}
\end{eqnarray}

\begin{eqnarray}\label{std1}
 \sigma  &=&  \sqrt{\frac{1}{N_{nucl}-1}\sum^{N_{nucl}}_{i=1}\left({x-\bar{x}}\right)^2}
\end{eqnarray}

In these equations, $x = Q_{Th}-Q_{Exp}$, $\bar{x}$ is the mean of $x$, and $N_{nucl}$ is the total number of data used for the calculations. The table ensuing exactitude of WS4 mass model in producing Q-values for all the considered decay modes. Hence, from here onwards, Q-values are taken from the WS4 mass model if the experimental Q-values are not available.\par
\begin{table}[!htbp]
\caption{RMSE and standard deviation ($\sigma$) of {Q$_{\alpha}$}, {Q$_{\beta^{-}}$}, and {Q$_{EC}$} for Pb isotopic chain.}
\centering
\resizebox{0.5\textwidth}{!}{%
%{\begin{tabular}{|c|c|c|c|c|c|c|c|c|c|c|c|c|c|c|c|}
{\begin{tabular}{@{}|l|ccc|ccc|}
 \hline
 \hline
  %\multicolumn{1}{|c|}{Nucleus}&
  \multicolumn{1}{|c|}{Theory}&
 \multicolumn{3}{c|}{RMSE}&
  \multicolumn{3}{c|}{$\sigma$}\\
 \cline{2-4} \cline{5-7}
  &{Q$_{\alpha}$}&{Q$_{\beta^{-}}$}& {Q$_{EC}$}&{Q$_{\alpha}$}&{Q$_{\beta^{-}}$}&  {Q$_{EC}$}\\
   \cline{1-4} \cline{5-7}
        WS4         &   0.36   &   0.29        &   0.53 &   0.06    &   0.13       &   0.56\\
        FRDM        &   0.37   &   0.32        &   0.29 &   0.09    &   0.18        &   0.30 \\
        TMA         &   0.61   &   1.48        &   0.63 &   0.27    &   0.50        &   0.85\\
        NL3*        &   1.06   &   1.89        &   1.70 &   0.37    &   0.23        &   0.08\\
        FSU-Garnet  &   1.13   &   2.10        &   1.27 &   0.28    &   0.47        &   0.58 \\
        FSU-Gold    &   1.17   &   3.37        &   1.91 &   1.10    &   0.73        &   0.36\\
        RCHB        &   2.06   &   0.62        &   0.79 &   1.35    &   0.23        &   0.31 \\
         \hline
   \hline
\end{tabular}}}
\label{rmse}
\end{table}

It is crucial to point out here that we have performed quadrupole constrained calculations using RMF approach for obtaining corresponding potential energy surfaces (PESs) in addition to the determination of the corresponding ground-state deformations for all the Pb isotopes. Most of the isotopes are found spherical or near spherical shapes, however, some neutron deficient nuclei $^{186,188}$Pb are found in coexisting oblate and prolate shapes. These Pb nuclei are already speculated
to show shape coexistence \cite{andreyev2000,duguet2003,hellemans2005}. The pronounced shape coexistence with prolate and oblate shapes appears to be a remarkable feature and characteristic of the nuclei away from shell closure that may have a significant role in the above mentioned Q-values and consequently in determination of lifetime\cite{crider2016}. The enhanced lifetime obtained with the second minima, which is lying at a very small energy difference may get favored in certain situations and affect the half-lives. Dependance of half-lives on deformed nuclei as well as nuclei having co-existence will be investigated in more detail in our future work.\par

\begin{table}[!t]
\caption{Coefficients of NMSF, NMHF, MTNF and QF formulas.}
%\centering
%\def\arraystretch{1.0}
\resizebox{1.00\textwidth}{!}{%
{\begin{tabular}{l|cccccccc}
%{\begin{tabular}{l|d@{\hskip 0.2in}d@{\hskip 0.2in}d@{\hskip 0.2in}d@{\hskip 0.2in}d@{\hskip 0.2in}d@{\hskip 0.2in}d@{\hskip 0.2in}c}
 \hline
\multicolumn{1}{c|}{Formulas}&
\multicolumn{1}{c}{a}&
 \multicolumn{1}{c}{b}&
\multicolumn{1}{c}{c}&
\multicolumn{1}{c}{d}&
 \multicolumn{1}{c}{e}&
 \multicolumn{1}{c}{f}&
  \multicolumn{1}{c}{g}&
   \multicolumn{1}{c}{$\overline{E}_{i}$}\\
\hline
     &             &       &  &  &  &                                                & & $\overline{E}_{i,ee}$ = 0.0000          \\
NMSF & 0.6975      &  -0.1113     & -27.1154 & 13.5610 & -24.7796 & 0.0174           &-& $\overline{E}_{i,eo}$ = -0.0396         \\
     &             &       &  &  &  &                                                & & $\overline{E}_{i,oe}$ = -0.0350         \\
     &             &       &  &  &  &                                                & & $\overline{E}_{i,oo}$ = 0.0438          \\
\hline

NMHF & 107.0131 & -206.5398         & -160.6152    & 309.6165 & 19.7237 & -31.1655 & 0.0238&- \\
\hline
MTNF & 0.7208      & -17.7056       & 0.0281     &  - & -&  - & -&-                   \\
\hline
QF & -0.3980      & 9.2318       & -75.3725     &  0.0354 & -&  - & -&-                   \\
\hline
\end{tabular}}}
\label{coefficients}
\end{table}
%------------------------------------------------------------------------------------------------------------------------------
\section{$\alpha$-decay}
One of the dominant decay modes in heavy and superheavy nuclei is the $\alpha$-decay for which energy release $Q_{\alpha}$ in ground-state to ground-state decay is obtained from mass excesses or total binding energies through
\begin{eqnarray}
Q_{\alpha} (MeV) & = & M(Z, N) - M(Z-2, N-2) - M(2, 2) \nonumber\\
& = & B.E.(Z-2, N-2) + B.E.(2, 2) - B.E.(Z, N)
\label{qalpha}
\end{eqnarray}
where the $^4He$ mass excess M(2, 2) is 2.42 MeV and the binding energy B.E.(2, 2) is 28.30 MeV. To calculate the $\alpha$-decay half-lives ($log_{10}T_{1/2}$), we have used four recently developed formulas. These formulas are fitted on latest experimental $\alpha$-decay half-lives, spin, and parity of 398 nuclei from NUBASE2020 \cite{audi20201} in the range 50$\leq$Z$\leq$118 which include an essential contribution of orbital angular momentum ($l$) of the emitted $\alpha$-particle and hence evidently described favoured and unfavoured $\alpha$-transitions \cite{sharmapk,saxenassj}. These formulas with brief details are mentioned below:

\subsection{New Modified Sobiczewski Formula (NMSF)}
Parkhomenkho and Sobiczewski have proposed the phenomenological formula to calculate the $\alpha$-decay half-lives \cite{Sobi2005} in 2005 which has been modified in our recent work \cite{Saxena2020}. In the present work, we use another modified version referred to as the new modified Sobiczewski formula (NMSF) \cite{sharmapk} which includes isospin dependency as well as the centrifugal term and given by:
\begin{eqnarray}
log_{10}T_{1/2}^{NMSF}(s) &=& aZ\sqrt{\mu}(Q_{\alpha} - \overline{E}_{i})^{-1/2} + bZ\sqrt{\mu} + c + dI \nonumber\\
&&+ eI^{2} + fl(l+1)
\label{nmsf}
\end{eqnarray}
where Z, $Q_{\alpha}$, $l$, and $\overline{E}_{i}$ represent the proton number, $\alpha$-decay energy of parent nucleus, minimum angular momentum and average excitation energy (in MeV), respectively. $\mu$ $=$ $\frac{A_d A_{\alpha}}{A_d + A_{\alpha}}$ represents reduced mass (where $A_d$ and $A_{\alpha}$ are mass number of the daughter nuclei and $\alpha$-particle, respectively), and I $=$ $\frac{N-Z}{A}$ exhibits nuclear isospin asymmetry. The values of fitted coefficients a, b, c, d, e, and f are given in Table \ref{coefficients}.

\subsection{New Modified Horoi Formula (NMHF)}\label{NMHF}
Another formula for calculating $\alpha$-decay half-lives along with cluster decay was offered by Horoi \textit{et al.} \cite{Horoi2004}. This formula has been modified in our recent work \cite{sharmapk} which depends upon the proton number of an alpha particle ($Z_{\alpha}$), proton number of daughter nucleus $Z_{d}$, $\alpha$-decay energy of parent nucleus ($Q_{\alpha}$), minimum angular momentum ($l$), reduced mass ($\mu$) and nuclear isospin asymmetry (I $=$ $\frac{N-Z}{A}$) given as:

\begin{eqnarray}
log_{10}T_{1/2}^{NMHF}(s) &=& (a\sqrt{\mu} + b)[(Z_{\alpha}Z_{d})^{0.6}Q_{\alpha}^{-1/2} - 7] + (c\sqrt{\mu} + d) + eI \nonumber \\
&&+ fI^{2} + gl(l+1)
\label{nmhf}
\end{eqnarray}
where the coefficients (a, b, c, d, e, f, and g) are given in Table \ref{coefficients}.

\subsection{Modified  TNF Formula (MTNF)}
One of the very old formulas i.e Tagepera-Nurmia formula \cite{TNF1961} has also been modified in one of our recent work \cite{saxenassj} and shapes as:

\begin{eqnarray}\label{mtnf}
log_{10}T_{1/2}^{MTNF}(s) &=& a\sqrt{\mu}(Z_d Q_{\alpha}^{-1/2}-Z_{d}^{2/3})+ b+cl(l+1)
\end{eqnarray}

Here $Z_{d}$, $Q_{\alpha}$, $\mu$, $l$ are same as defined above, and the fitted coefficients (a, b, and c) are given in Table \ref{coefficients}.
\subsection{Quadratic Fitting Formula (QF)}
An entirely new formula was also proposed \cite{saxenassj} with only 4 fitted coefficients incorporating a quadratic fitting term. The formula is named as QF formula and given by:
 \begin{eqnarray}\label{SSJ}
log_{10}T_{1/2}^{QF}(s) &=& a\sqrt{\mu}\left(\frac{Z_d^{0.6}}{\sqrt{Q_{\alpha}}}\right)^2 + b\sqrt{\mu}\left(\frac{Z_d^{0.6}}{\sqrt{Q_{\alpha}}}\right)
+c+dl(l+1)
\end{eqnarray}
where the symbols are defined above and the coefficients (a, b, c, and d) are given in Table \ref{coefficients}.

\begin{table*}[!htbp]
\caption{$\alpha$-decay half-lives for Pb isotopes. Experimental data are taken from Refs. \cite{audi20202,audi20201}.}
\centering
\resizebox{0.8\textwidth}{!}{%
%{\begin{tabular}{|c|c|c|c|c|c|c|c|c|c|c|c|c|c|c|c|}
{\begin{tabular}{|c|c@{\hskip 0.2in}c@{\hskip 0.2in}c@{\hskip 0.2in}c|c@{\hskip 0.2in}c@{\hskip 0.2in}c@{\hskip 0.2in}c|}
 \hline
 \hline
  \multicolumn{1}{|c|}{}&
  \multicolumn{4}{|c|}{Exp.}&
  %\multicolumn{1}{c|}{Theory}&
 \multicolumn{4}{|c|}{{$log_{10}T_{1/2}^{cal}$} (s)}\\
 \cline{2-5}
 \cline{6-9}
  {A}&{$Q_\alpha$}&$l_{min}$ &$log_{10} T_{1/2}$  &Decay modes&{NMSF}&{NMHF}&{MTNF}&{QF}\\
  &(MeV)& &(s)  &&&&&\\
 \hline
 176  &7.88*&0 &  -   &  -                                        &-4.06  &-4.29  &-3.54 &-4.42        \\
 177  &7.69*&3 &  -   &  -                                        &-3.42   &-3.49  &-2.72 &-3.37                \\
 178  &7.79 &0 &-3.92 &$\alpha$                                   &-3.73   &-3.95  &-3.31 &-4.12                 \\
 179  &7.60 &2 &-2.46 &$\alpha$                                   &-3.17   &-3.26  &-2.63 &-3.25                            \\
 180  &7.42 &0 &-2.39 &$\alpha$                                   &-2.63   &-2.88  &-2.30 &-2.84                    \\
 181  &7.24 &2 &-1.44 &$\alpha$                                   &-2.08   &-2.19  &-1.62 &-1.99                         \\
 182  &7.07 &0 &-1.26 &$\alpha$                                   &-1.51   &-1.79  &-1.27 &-1.56                             \\
 183  &6.93 &2 &-0.27 &$\alpha$                                   &-1.06   &-1.18  &-0.67 &-0.82               \\
 184  &6.77 &0 &-0.31 &$\alpha$                                   &-0.52   &-0.80  &-0.35 &-0.44              \\
 185  &6.70 &2 &0.80  &$\alpha$/$\beta^+/EC$                      &-0.25   &-0.36  &0.08  &0.09                          \\
 186  &6.47 &0 &0.68  &$\beta^+/EC$/$\alpha$                      &0.57   &0.28   &0.67  &0.78  \\
 187  &6.39 &2 &1.18  &$\beta^+/EC$/$\alpha$                      &0.84   &0.73   &1.11  &1.31                                  \\
 188  &6.11 &0 &1.40  & $\beta^+/EC$/$\alpha$                     &1.94   &1.63   &1.98  &2.31              \\
 189  &5.92 &2 &1.59  & $\beta^+/EC$/$\alpha$                     &2.67   &2.55   &2.90  &3.38           \\
 190  &5.70 &0 &1.85  & $\beta^+/EC$/$\alpha$                    &3.63   &3.31   &3.62  &4.16          \\
 191  &5.46 &0 &1.90  & $\beta^+/EC$/$\alpha$                     &4.51   &4.36   &4.65  &5.29       \\
 192  &5.22 &0 &2.32  & $\beta^+/EC$/$\alpha$                     &5.81   &5.48   &5.75  &6.48           \\
 193  &5.01 &0 &2.54  & $\beta^+/EC$                              &6.67   &6.54   &6.80  &7.57                      \\
 194  &4.74 &0 &2.81  & $\beta^+/EC$                              &8.33   &8.00   &8.24  &9.05                    \\
 204  &1.97 &0 &$\geq24.64$   & $\alpha$                          &37.26  &37.12  &37.19 &28.32               \\ \hline
 \hline
\end{tabular}}}
\label{alpha-decay}
\end{table*}

Using these 4 formulas, the half-lives for the ground to ground state decay have been calculated for $^{176-194}$Pb and $^{204}$Pb isotopes that are given in Table \ref{alpha-decay}. For these half-lives, the minimum angular momentum $l_{min}$ (written as $l$ in the formulas) taken away by $\alpha$-particle is obtained by using standard selection rules \cite{denisov2009} based on spin and parity of parent and daughter nuclei. For comparison, experimental half-lives along with probable decay modes are also mentioned \cite{audi20201}. From Table \ref{alpha-decay}, it is indulging to note that our formulas are adequate to estimate $\alpha$-decay half-lives and are in a close match with the experimental half-lives where the probability of $\alpha$-decay is found dominant. Towards the neutron deficient side, we have estimated half-lives of $^{176-177}$Pb for which Q-values are taken from the WS4 mass model (shown by an asterisk (*)) as concluded in the previous section. It is important to note that the NMHF formula is found to be more accurate with minimum RMSE value among the considered formulas for Pb isotopes, specifically.\par

The study of Pb isotopes provides an excellent ground where the $\alpha$-decay competes with $\beta^+/EC$-decay towards neutron deficient side as can also be seen from Table \ref{alpha-decay} for the $^{185-192}$Pb. With this in view, we have estimated half-lives of weak-decay for isotopes of Pb having A$>$184. The process to calculate half-life for weak-decay ($\beta^{-}$ \& $\beta^{+}/EC$) is elaborated in the next section.

%---------------------------------------------------------------------------------------------------------------------
\section{$\beta$-decay \& Electron Capture (Weak-decay)}
The energy released in ground-state to ground-state electron-decay ($\beta^-$-decay) and electron capture (EC) are given in terms of the atomic mass excess $M(Z,N)$ or the total binding energy $B.E.(Z,N)$ by:

\begin{eqnarray}
        %\begin{aligned}[b]
Q_{\beta^-} (MeV) & = &  M(Z, N) - M(Z+1, N-1) \nonumber\\
& = &B.E.(Z+1, N-1) - B.E.(Z, N) + M_n -M _H
       % \end{aligned}
       \label{qbetaminus}
\end{eqnarray}
\begin{eqnarray}
Q_{EC} (MeV)&=&M(Z, N) - M(Z-1, N+1) - \mbox{B.E$^{e^{-}}$} \nonumber \\
&=&B.E.(Z-1, N+1) - B.E.(Z, N) + M_H -M_n-\mbox{B.E$^{e^{-}}$}
         \label{qec}
\end{eqnarray}

To look into the possibility of weak-decay, we adopt the empirical formula of Fiset and Nix \cite{Fiset1972} for estimating the corresponding half-lives. It is worthy to note that this formula of $\beta$-decay has recently been used successfully in a few of our works \cite{saxenaijmpe2019,saxena4} and the work by Ikram \textit{et al.} \cite{Ikram2017}.

        %\begin{aligned}[b]
        \begin{equation}\label{tbeta}
         T_{\beta} (s) = \frac{540 m_e^5}{\rho(W_\beta^6-m_e^6)}\times 10^{5.0}
        \end{equation}

This equation (\ref{tbeta}) is only logical for $W_{\beta}\gg m_e$. For the average density of states $\rho$ in the daughter nucleus, we use the empirical results given by Seeger \textit{et al.} \cite{Seeger65}, from which $\rho$ has values 0.97 and 1.67 for even and odd A isotopes of Pb, respectively. The formula for electron capture (EC) is given by
\begin{eqnarray}
 T_{EC} (s) &= &\frac{9 m_e^2}{2\pi(\alpha Z_K)^{2s+1}\rho\left[Q_{EC}-(1-s)m_e\right]^3} \left(\frac{2R_0}{\hbar c/ m_e}\right)^{2-2s} \nonumber\\
 &&\times\frac{\Gamma(2s+1)}{1+s}\times10^{6.5}
             \label{tecfinal}
\end{eqnarray}

Here, $Z_K$ is the effective charge of the parent nucleus for an electron in the K-shell; it is given approximately by $Z_K= Z_P - 0.35$, The energy $W_{\beta}$ is sum of energy of the emitted $\beta-particle$ and its rest mass $m_e$. i.e. $W_{\beta} = Q_{\beta}+m_{e}$. Also, the quantity s is given by $s = [ 1 -(\alpha Z_K)^2]^{\frac{1}{2}}$ and represents the rest mass of an electron minus its binding energy in the K-shell, in units of $m_e$. The quantity $\alpha$ is the fine-structure constant, and $R_0$ is the nuclear radius, which is taken to be $R_0 = 1.2249 A^{\frac{1}{3}} fm$.\par

In this paper, we follow Eqn.(\ref{tbeta}) to calculate half-lives for $\beta^-$-decay whereas the Eqn.(\ref{tecfinal}) is used to calculate half-lives for electron capture (EC). The Table \ref{betaEc} consists of half-lives of $^{185-203}$ Pb calculated for $\beta^+$/EC-decay. In a similar manner, $\beta^-$-decay half-lives are estimated for a wide range which are mentioned in Table \ref{betaminus} for $^{209-266}$Pb. The experimental half-lives using the empirical formulas are in reasonable match with experimental data, which automatically validates the use of the formula in the unknown region. As mentioned above, the majority of the Q-values are taken from the WS4 mass model for $\beta^-$-decay.\par

\par
\begin{table*}[!ht]
\caption{ $\beta^+/EC$-decay half-lives for Pb isotopes. Experimental data are taken from Refs. \cite{audi20202,audi20201}.}
\centering
%\def\arraystretch{1.0}
%\resizebox{0.4\textwidth}{!}{%
\resizebox{1.00\textwidth}{!}{%
%{\begin{tabular}{|c|c|c|c|c|c|c|c|c|c|c|c|c|c|c|c|}
{\begin{tabular}{|c c@{\hskip 0.2in}c@{\hskip 0.2in} c@{\hskip 0.2in}c@{\hskip 0.2in}c||cc@{\hskip 0.2in}c@{\hskip 0.2in} c@{\hskip 0.2in}cc|c@{\hskip 0.2in}c@{\hskip 0.2in}c|c@{\hskip 0.2in}||c|c|c@{\hskip 0.2in}c@{\hskip 0.2in}c|c@{\hskip 0.2in}c|c|c@{\hskip 0.2in}c@{\hskip 0.2in}cc@{\hskip 0.2in}c|c|c@{\hskip 0.2in}c@{\hskip 0.2in}cc@{\hskip 0.2in}c}
%|c|c@{\hskip 0.2in}c@{\hskip 0.2in}c|c@{\hskip 0.2in}c}
%||c|c@{\hskip 0.2in}c@{\hskip 0.2in}c|c@{\hskip 0.2in}c|c|c@{\hskip 0.2in}c@{\hskip 0.2in}c|c@{\hskip 0.2in}c}
 \hline
 \hline
  \multicolumn{1}{|c}{}&
  \multicolumn{3}{c}{Exp.}&
   \multicolumn{1}{c}{}&
   \multicolumn{1}{c||}{{$log_{10}T_{1/2}^{cal}$} (s)}&
   \multicolumn{1}{c}{}&
  \multicolumn{3}{c}{Exp.}&
  \multicolumn{1}{c}{}&
   \multicolumn{1}{c|}{{$log_{10}T_{1/2}^{cal}$} (s)} % \multicolumn{1}{|c|}{}&
  %\multicolumn{3}{c|}{Exp.}&
   %\multicolumn{2}{c|}{{$log_{10}T_{1/2}^{cal}$}}

 \\
 \cline{2-4}
 \cline{6-6}
 \cline{8-10}
 \cline{12-12}
 %\cline{14-18}
  {A}&{$Q_{EC}$}&{$log_{10}T_{1/2}$}  &{Decay modes}&&{$\beta^+$}/{EC} &  {A}&{$Q_{EC}$}&{$log_{10}T_{1/2}$}  &{Decay modes}&&{$\beta^+$}/{EC}\\
  &(MeV)&(s)&&& &  &(MeV)&(s)&&&\\
  %&  {A}&{$Q_{\beta^+}$}&{$log_{10}T_{1/2}$}  &{Decay modes}&{$\beta^+$}&{EC}\\
 \hline
 185  &8.22  &0.80  &$\alpha$/$\beta^+/EC$      &    &3.04 &194  &2.73 &2.81  & $\beta^+/EC$              &     &4.75      \\
 186  &5.20  &0.68  &$\beta^+/EC$/$\alpha$      &    &3.88 &195  &4.45 &2.95  &$\beta^+/EC$               &     &3.86            \\
 187  &7.46  &1.18  &$\beta^+/EC$/$\alpha$      &    &3.17 &196  &2.15  &3.35 &$\beta^+/EC$               &     &5.07               \\
 188  &4.52  &1.40  & $\beta^+/EC$/$\alpha$     &    &4.07 &197  &3.60 &2.69   &$\beta^+/EC$              &     &4.14       \\                                                                                                                                                                                                               189  &6.77  &1.59  & $\beta^+/EC$/$\alpha$     &    &3.30 &198  &1.46 &3.94   &$\beta^+/EC$              &     &5.61           \\
 190  &3.96  &1.85  & $\beta^+/EC$/$\alpha$    &    &4.25 &199  &2.83 &3.73   &$\beta^+/EC$              &     &4.46                \\
 191  &6.05  &1.90  & $\beta^+/EC$/$\alpha$     &    &3.45 & 200 &0.80 &4.89   &$\beta^+/EC$             &      &6.48            \\
 192  &3.32 &2.32  & $\beta^+/EC$/$\alpha$      &    &4.48 & 201 &1.91 &4.53   &$\beta^+/EC$             &      &5.00           \\
 193  &5.28  &2.54  & $\beta^+/EC$              &    &3.63 & 203 &0.98 &5.27   &$\beta^+/EC$             &      &5.95               \\
 \hline
\hline
\end{tabular}}}
\label{betaEc}
\end{table*}

%------------------------------------------------------------------------------------------------------------------------------------------------------
\begin{table*}[!htbp]
%\begin{table*}[hbt!]
\caption{$\beta^-$-decay half-lives for Pb isotopes. Experimental data are taken from Refs. \cite{audi20202,audi20201}. Other Q-values are taken from WS4 mass model (kindly refer to section \ref{section-theory}).}
\centering
\resizebox{1.0\textwidth}{!}{%
%{\begin{tabular}{|c|c|c|c|c|c|c|c|c|c|c|c|c|c|c|c|}
{\begin{tabular}{|c@{\hskip 0.2in}c@{\hskip 0.2in}c@{\hskip 0.2in}c@{\hskip 0.2in}c||c@{\hskip 0.2in}c@{\hskip 0.2in}c||c@{\hskip 0.2in}c@{\hskip 0.2in}c||c@{\hskip 0.2in}c@{\hskip 0.2in}c|}
 \hline
 \hline
   \multicolumn{1}{|c}{}&
  \multicolumn{3}{c}{Exp.}&
 \multicolumn{1}{c||}{{$log_{10}T_{1/2}^{cal}$}} &
   \multicolumn{1}{c}{}&
   \multicolumn{1}{c}{WS4}&
  \multicolumn{1}{c||}{{$log_{10}T_{1/2}^{cal}$}}&
   \multicolumn{1}{c}{}&
  \multicolumn{1}{c}{WS4}&
 \multicolumn{1}{c||}{{$log_{10}T_{1/2}^{cal}$}} &
   \multicolumn{1}{c}{}&
   \multicolumn{1}{c}{WS4}&
  \multicolumn{1}{c||}{{$log_{10}T_{1/2}^{cal}$}}\\
 \cline{2-4}
 \cline{7-7}
 \cline{10-10}
 \cline{13-13}
{A}&{$Q_{\beta^-}$}&{$log_{10}T_{1/2}$}  &{Decay modes}&{($\beta^-$)}& {A}&{$Q_{\beta^-}$}&{($\beta^-$)}&{A}&{$Q_{\beta^-}$}&{$(\beta^-$)}& {A}&{$Q_{\beta^-}$}&{($\beta^-$)}\\
&(MeV)&(s)&&(s)& &(MeV)&(s)&&(MeV)&(s)& &(MeV)&(s)\\ \hline
 209  &0.64   &4.07        &$\beta^-$  &5.68 &221  &5.21      &1.51 &237  &10.49    &-0.20  &253  &14.48    &-1.00      \\
 210  &0.06   &8.80        &$\beta^-$  &8.03 &222  &4.43      &2.13 &238  &9.62     &0.25   &254  &13.82    &-0.65       \\
 211  &1.37   &3.34        &$\beta^-$  &4.41 &223  &6.36      &1.03 &239  &11.02    &-0.32  &255  &15.20    &-1.12       \\
 212  &0.57   &4.58        &$\beta^-$  &6.09 &224  &5.45      &1.64 &240  &10.12    &0.13   &256  &14.49    &-0.77       \\
 213  &2.03   &2.79        &$\beta^-$  &3.62 &225  &7.03      &0.79 &241  &11.50    &-0.43  &257  &15.68    &-1.20       \\
 214  &1.02   &3.21        &$\beta^-$  &5.18 &226  &6.04      &1.39 &242  &10.65    &0.00   &258  &14.95    &-0.85       \\
 215  &2.71   &2.17        &$\beta^-$  &3.00 &227  &7.56      &0.61 &243  &12.09    &-0.55  &259  &16.10    &-1.27       \\
 216  &1.61   &$>-6.52$    &$\beta^-$  &4.33 &228  &6.53      &1.20 &244  &11.11    &-0.10  &260  &15.44    &-0.93       \\
 217  &3.50   &$>-6.52$    &$\beta^-$  &2.43 &229  &8.08      &0.45 &245  &12.00    &-0.53  &261  &16.68    &-1.36       \\
 218  &2.20   &$>-6.52$    &$\beta^-$  &3.69 &230  &7.08      &1.00 &246  &10.79    &-0.03  &262  &15.92    &-1.01       \\
 219  &4.00   &$>-6.52$    &$\beta^-$  &2.13 &231  &8.57      &0.30 &247  &12.47    &-0.63  &263  &17.04    &-1.41       \\
 220  &2.90   &$>-6.52$    &$\beta^-$  &3.09 &232  &7.73      &0.79 &248  &11.82    &-0.26  &264  &16.37    &-1.08       \\
 &      &           &          &             &233  &9.25      &0.11 &249  &13.29    &-0.79  &265  &17.44    &-1.47       \\
 &      &           &          &             &234  &8.37      &0.60 &250  &12.49    &-0.40  &266  &16.88    &-1.15       \\
 &      &           &          &             &235  &9.84      &-0.04&251  &13.83    &-0.89   &&&\\
  &      &           &          &            &236  &9.01      &0.41 &252  &13.15    &-0.53   &&&\\
\hline
 \hline
\end{tabular}}}
\label{betaminus}
\end{table*}
%\FloatBarrier

%-------------------------------------------------------------------------------------------------------------------------------------------------

%------------------------------------------------------------------------- -------------------------------------------------------------------------
\begin{figure}[!htbp]
\centering
\includegraphics[width=0.8\textwidth]{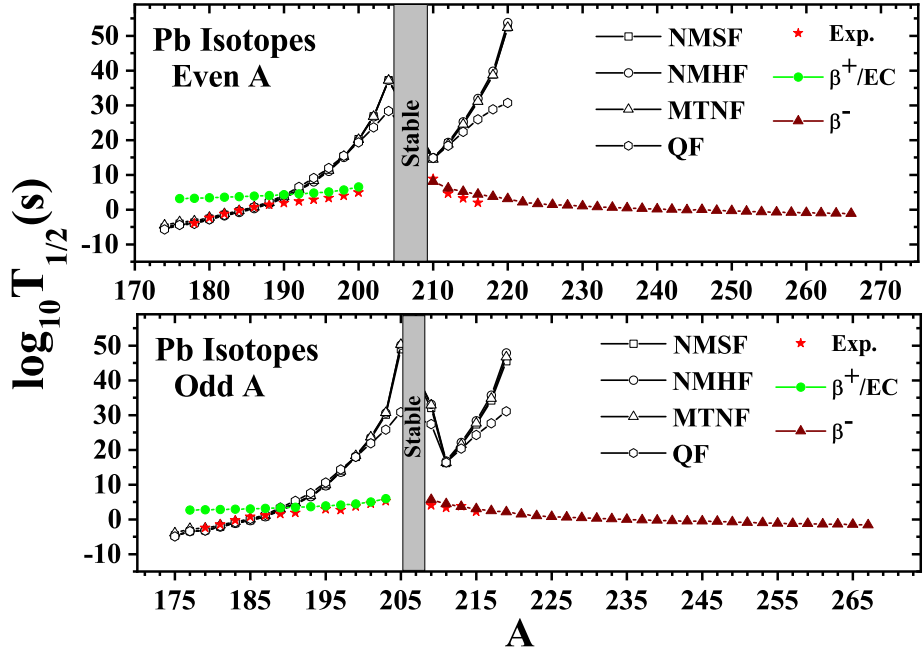}
\caption{(Colour online) The comparison of half-lives of $\alpha$, $\beta^{-}$, $\beta^{+}$/EC-decays.}\label{fig1}
\end{figure}

To visualize the competition among all these concerned decay modes and their agreement with available experimental half-lives, we have plotted our calculated half-lives for $\alpha$, $\beta^{-}$-decays as well as electron capture (EC) in Fig. \ref{fig1} along with experimental half-lives \cite{audi20201}. An excellent match with experimental data renders our estimation of half-lives with a good accuracy which is adequate to use in all regions of the periodic chart. A close look of Fig. \ref{fig1} suggests shell effect near N$=$126 as the $\alpha$-decay half-lives approach to very high values. This region is also marked as 'stable' in the figure. The competition among considered decays demonstrates a clear possibility of $\alpha$-decay in the neutron deficient region which is subsequently overtaken by $\beta^{+}$/EC-decay upto the stable region. After stable region, $\beta^{+}$/EC-decay is energetically forbidden, however, $\beta^{-}$-decay becomes more probable and holds up to the neutron drip-line.\par
%-------------------------------------------------------------------------------------------------
To examine the competition among various decay modes, we have calculated the total half-lives of Pb isotopes in the range 103$\leq$N$\leq$117 (using Eqn. \ref{THL}), and compared them with the available experimental data.\par

\begin{equation}\label{THL}
  \frac{1}{T^{Th.}_{1/2}} = \frac{1}{T^{\alpha}_{1/2}}+\frac{1}{T^{\beta^+/EC}_{1/2}}
\end{equation}
%\frac{1}{T^{Th.}_{1/2}} &=& \frac{1}{T^{\alpha}_{1/2}}+\frac{1}{T^{\beta^+}_{1/2}}+\frac{1}{T^{EC}_{1/2}}
%\label{THL}

where $\alpha$-decay half-lives (T$^{\alpha}_{1/2}$) have calculated by using NMHF (as mentioned in subsection \ref{NMHF}), $\beta^+$/EC-decay half-lives calculated with the help of Eqn. (\ref{tecfinal}). Thereafter, to quantize the competition among various decay-modes, we have calculated the branching ratios for respective decay modes as:
\begin{eqnarray}
   b=\left\{
    \begin{array}{ll}
       \frac{T^{Th.}_{1/2}}{T^{\alpha}_{1/2}}\,\,\,\,\,\,
       &\mbox{for $\alpha$-decay}\\
       \frac{T^{Th.}_{1/2}}{T^{\beta^+/EC}_{1/2}}\,\,\,\,\,\,
       &\mbox{for $\beta^+$/EC-decay}\\

      \end{array}\right.
      \label{selection-rules}
\end{eqnarray}

%\begin{equation}\label{branch}
%  b  = \frac{T^{Th.}_{1/2}}{T^{\alpha/\beta^+/EC}_{1/2}}
%\end{equation}
The values of total half-lives and branching ratios are listed in Table \ref{branching}. It is to be noted that the nuclei with A$\leq$184 own $\alpha$-decay as the only dominant mode. Similarly, nuclei with A$\geq$209 have only $\beta^{-}$-decay mode predominantly. For the rest of the nuclei, the competition is clearly evident from the Table \ref{branching} where the probability of $\alpha$-decay decreases gradually towards neutron-rich side and chances of $\beta^{+}$/EC-decay turn more probable. It is also satisfactory to note that total half-lives calculated by using Eqn. (\ref{THL}) is in a reasonable match with the experimental data.
\begin{table*}[!htbp]
\caption{$\alpha$-decay total half-lives and branching ratio for Pb isotopes. Experimental data are taken from Ref. \cite{audi20201}.}
\centering
\footnotesize
\resizebox{1.00\textwidth}{!}{%
%{\begin{tabular}{|c|c|c|c|c|c|c|c|c|c|c|c|c|c|c|c|}
{\begin{tabular}{|cc@{\hskip 0.2in}c@{\hskip 0.2in}c@{\hskip 0.2in}c@{\hskip 0.2in}c||c@{\hskip 0.2in}c@{\hskip 0.2in}cc@{\hskip 0.2in}c@{\hskip 0.2in}c|}
 \hline
 \hline
  \multicolumn{1}{|c}{}&
   %\multicolumn{1}{|c|}{}&
   % \multicolumn{1}{|c|}{}&
  \multicolumn{2}{c}{Total $log_{10}T_{1/2}$ (s)}&
  \multicolumn{1}{c}{}&
  %\multicolumn{1}{c|}{Theory}&
 \multicolumn{2}{c||}{{Branching ratio}}&
  \multicolumn{1}{c}{}&
   %\multicolumn{1}{|c|}{}&
   % \multicolumn{1}{|c|}{}&
  \multicolumn{2}{c}{Total $log_{10}T_{1/2}$ (s)}&
   \multicolumn{1}{c}{}&
  %\multicolumn{1}{c|}{Theory}&
 \multicolumn{2}{c|}{{Branching ratio}}\\
 \cline{2-3}
 \cline{5-6}
 \cline{8-9}
 \cline{11-12}
  {A}&Exp.  &Theoretical&&{$\alpha$}&{$\beta^+$}/{EC}&{A}&Exp.  &Theoretical&&{$\alpha$}&{$\beta^+$}/{EC}\\
 \hline
 185  &0.80&-0.36   &&99.96    &0.04    &194&2.81 &4.51   &&0.03     &99.97      \\
 186  &0.68&0.28    &&99.98    &0.02    &195&2.95 &4.09   &&0.00     &100.00        \\
 187  &1.18&0.73    &&99.64    &0.36    &196&3.35 &4.84   &&0.00     &100.00        \\
 188  &1.40&1.63    &&99.37    &0.63    &197&2.69 &4.38   &&0.00     &100.00    \\
 189  &1.59&2.51    &&90.53    &9.47    &198&3.94 &5.37   &&0.00     &100.00    \\
 190  &1.85&3.23    &&83.30    &16.70   &199&3.73 &4.70   &&0.00     &100.00    \\
 191  &1.90&3.60    &&17.34    &82.66   &200&4.89 &6.25   &&0.00     &100.00    \\
 192  &2.32&4.22    &&5.50     &94.50   &201&4.53 &5.24   &&0.00     &100.00     \\
 193  &2.54&3.86    &&0.21     &99.79   &203&5.27 &6.18   &&0.00     &100.00      \\
  \hline
 \hline
\end{tabular}}}
\label{branching}
\end{table*}

\section{Cluster Decay}
Another exotic decay, in which a heavy nucleus decays into a light daughter nucleus and emits a fragment, is named cluster-decay. This kind of decay was first studied by Sandulescu \textit{et al.} in 1980 \cite{Sandulescu} which is subsequently observed experimentally \cite{Rose} in $^{223}$Ra which decays into a large fragment near doubly-magic $^{208}$Pb and a lighter cluster $^{14}$C. Many of the experiments have been done in this region \cite{Aleksandrov1984,Gales1984,Kutschera1985} and several other clusters like $^{20}$O, $^{22,24-26}$Ne, $^{28-30}$Mg and $^{32,34}$Si have been experimentally observed so far \cite{Barwick1986,Bonetti1993,Bonetti1994} which predominantly diverge towards Pb isotopes. Hence the emission of the cluster is closely related to the Pb isotopes in the heavy and superheavy regions which incited our investigation towards cluster emission. At this point, it is clearly mentioned that unlikely to above-considered decay modes, the probability of cluster emission from Pb isotopes is insignificant or negligible, however, Pb isotopes own their candidature to become daughter nuclei in cluster emission of heavier and superheavy nuclei. With this in view, considering daughter nuclei as Pb nuclei the cluster emission is looked in for heavy and superheavy nuclei. To calculate the energy released in cluster decay, we have used the following formula:

\begin{figure*}[!t]
\centering
\includegraphics[width=1.00\textwidth]{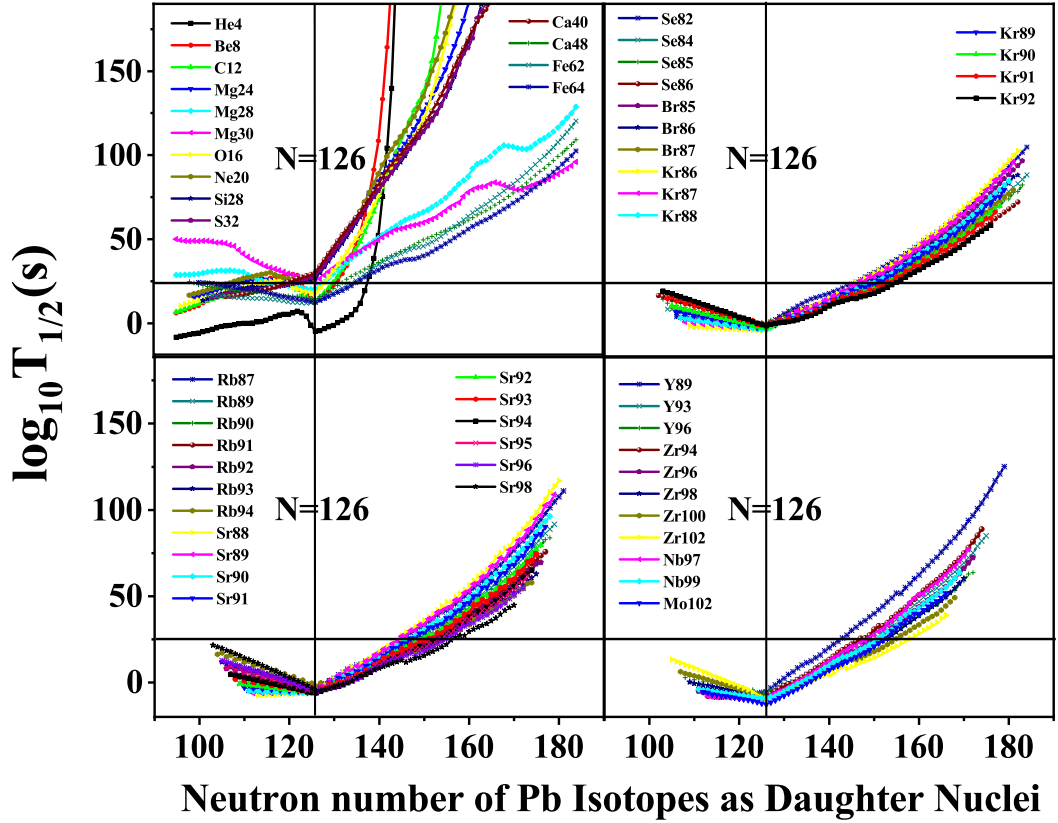}
\caption{(Colour online) Half-lives for various cluster emission considering Pb isotopes as daughter nuclei.}\label{fig2}
\end{figure*}

\begin{equation}\label{Qc}
  Q (MeV) = B.E.(d) + B.E.(c) - B.E.(p)
\end{equation}

where B.E.(d), B.E.(c), and B.E.(p) are the binding energies of the daughter, cluster, and parent nucleus, respectively and taken from WS4 mass table \cite{ws4}. To estimate half-life of cluster decay, a very widely known formula proposed by Qi \textit{et al.} \cite{UDL} is used which is also called universal decay law (UDL) and given as:
\begin{eqnarray}
log_{10}T_{1/2}^{UDL}(s) &=& aZ_{c}Z_{d}\sqrt{\frac{\mu}{Q}} + b[\mu Z_{c}Z_{d}({A_{c}}^{1/3}
+ {A_{d}}^{1/3})]^{1/2} + c
\label{equdl}
\end{eqnarray}
where $\mu$ is the reduced mass as mentioned above, and $Z_{c}$, $Z_{d}$, Q are proton number of cluster nucleus, proton number of daughter nucleus, the energy released during decay, respectively. Also, a$=$0.3949, b$=$-0.3693, and c$=$-23.7615 are fitting coefficients. The formula is applied for a wide range of Pb nuclei considered as daughter nuclei and covering clusters from $^{4}$He to $^{102}$Mo to calculate half-lives, which is plotted in Fig. \ref{fig2}. The strong stability of $^{208}$Pb clearly visible in all the panels of Fig. \ref{fig2} as all the half-lives are found minimum for all the considered clusters. This investigation evidently establishes cluster decay as one of the important decay modes in heavy and superheavy nuclei which needs separate investigation.
\section{Conclusions}
Various decay modes are probed on equal footing which are mainly concerned with Pb isotopes. Towards the neutron deficient side, half-lives are estimated using four newly proposed empirical formulas of $\alpha$-decay (NMSF, NMHF, MTNF, and QF). Similarly, half-lives towards both sides of the valley of $\beta$-stability are calculated using the empirical formula of weak-decay. Calculated half-lives along with the competition of these decay modes are found consistent with experimental values as a result these formulas are utilized to estimate half-lives for other unknown isotopes of Pb. It is concluded that for the Pb isotopes the WS4 mass table results more accurately for calculating Q-values of decay modes. In particular, for $\alpha$-decay, the NMHF formula is found to be more precise for Pb isotopes and hence can be used to estimate $\alpha$-decay half-lives of unknown nuclei in this region of the periodic chart. Another important outcome of our study is to bring in the possibility of cluster emission from superheavy nuclei which produces Pb isotopes as one of the daughter nuclei by emitting clusters ranging from He to Mo isotopes. This comprehensive study of decay modes linked with Pb isotopes is expected to provide a reasonable drive to the experiments eyeing on Pb or heavier nuclei.
\section{Acknowledgement}
The support provided by SERB (DST), Govt. of India under CRG/2019/001851 is acknowledged.
%%%%%%%%%%%%%%%%%%%%%%%%%%%%%%%%%%%%%%%%%%%%%%%%%%%%%%%%%%%%%%%%%%%%%%%%%%%%%%%%%%%%%%%%%%%%%%%%%%%%%%%%%%%%%%%%%%%%%%%%%%%%%%%%%%%%%%%%%%%%%%

\end{document}